# Highlights

**An explainable hybrid deep learning-enabled intelligent fault detection and diagnosis approach for automotive software systems validation**

Mohammad Abboush, Ehab Ghannoum, Andreas Rausch

- Proposing an intelligent test recordings analysis approach of automotive software systems (ASSs) based on explainable deep learning methods.

- Developing a hybrid 1dCNN-GRU-based DL model architecture for fault detection, identification, and localization considering single and concurrent faults with imbalanced data.

- Providing a white box DL model version with identification of significant variables based on interpretation of prediction results improving performance and efficiency along with reducing complexity.

- Investigating and analyzing four explainable AI techniques, highlighting their different computational costs and performance results.

- Employing HIL real-time simulation dataset to validate the proposed approach as an industrial application, considering user behavior and the high-fidelity ASSs models.

# An explainable hybrid deep learning-enabled intelligent fault detection and diagnosis approach for automotive software systems validation

Mohammad Abboush*, Ehab Ghannoum and Andreas Rausch·

*Technische Universität Clausthal, Institute for Software and Systems Engineering, Arnold-Sommerfeld-Straße 1, Clausthal-Zellerfeld, 38678, Niedersachsen, Germany*

## ABSTRACT

Advancements in data-driven machine learning have emerged as a pivotal element in supporting automotive software systems (ASSs) engineering across various levels of the V-development process. During system verification and validation, the integration of an intelligent fault detection and diagnosis (FDD) model with test recordings analysis process serves as a powerful tool for efficiency ensuring functional safety. However, the lack of interpretability of the black-box FDD models developed not only hinders understanding of the cause underlying the prediction, but also prevents the model from being adapted based on the prediction result. This, in turn, increases the computational cost required for developing a complex FDD model and limits confidence in real-time safety-critical applications. To address this challenge, a novel explainable method for fault detection, identification, and localization is proposed in this article with the aim of providing a clear understanding of the logic behind the prediction outcome. To this end, a hybrid 1dCNN-GRU-based intelligent model was developed to analyze the recordings from the real-time validation process of ASSs. The employment of explainable AI techniques, i.e., IGs, DeepLIFT, Gradient SHAP, and DeepLIFT SHAP, was instrumental in enabling model adaptation and facilitating the root cause analysis (RCA). The proposed approach is applied to the real-time dataset collected during a virtual test drive performed by the user on hardware-in-the-loop (HIL) system. A thorough evaluation of the model's performance revealed its superiority in diagnosing the fault type and location compared to state-of-the-art models. By integrating XAI techniques, the proposed approach offers a high-performance, low-computational-cost white-box DL model, which not only supports safety engineers but also contributes to the optimization of real-time validation process of ASSs.

## 1. Introduction

In the automotive domain, the system development life cycle encompasses exhaustive testing activities that are conducted in accordance with the ISO 26262 standard [1]. Depending on the state of the system under test (SUT), the aforementioned activities are performed at different phases of the V-model development, known as X-in-the-Loop tests[18]. In the concluding phase preceding the commencement of production, a virtual/ real test drive is conducted to validate the real vehicle systems against the requirements [30, 26].The utilization of hardware-in-the-loop (HIL) simulations alongside physical vehicle components and actual electronic control units (ECUs) allows for the efficient conduct of real-time validation process. [27]. However, the complexity of the recorded system behaviour, involving a substantial amount of data, poses a challenges for the implementation of conventional knowledge-based analysis methods and increases the associated processing costs [47, 23]. Current approaches employing industry-certified tools encounter obstacles in identifying the root cause of faults and distinguishing between safe and critical faults [11]. In view of the challenges outlined above, it is crucial to develop an efficient approach for analysing test recordings during the real-time validation process of ASSs.

Recently, a data-driven approach has attracted researchers' attention to investigate the development of an advanced fault detection and diagnosis (FDD) model based on historical datasets [38]. The primary rationales for the superiority of this approach over alternative approaches, including knowledge-based [12], model-based [22], and signal-based approaches [17], are its low cost, ease of use, high performance, and flexibility. In various domains, the integration of data-driven machine learning (ML) and deep learning (DL) techniques into the development process of FDD model has proven to be a promising way to overcome various engineering challenges [39]. In the automotive industry, for example, it is noteworthy that the application of the DL-based FDD model for analyzing vehicle test reports contributes to reducing the costs of the validation process in terms of time, effort, and difficulty [49]. In this manner, valuable insights into the nature and causes of faults during the testing process can be obtained in an optimal manner. However, due to the "black box" characteristic of the developed FDD models, the decision-making behind the diagnosis process is invisible to engineers [50]. Consequently, the process of FDD model optimisation and adaption is rendered more complex. Further challenges arise from the complexity of faulty data patterns, such as the occurrence of simultaneous faults [10], which poses a significant obstacle to explain the diagnostic process [56]. This emphasises a gap in the current-state-of the-art methods concerning the

---
*Corresponding author

✉ mohammad.abboush@tu-clausthal.de (M. Abboush);
ehab.ghannoum@tu-clausthal.de (E. Ghannoum);
andreas.rausch@tu-clausthal.de (A. Rausch)

🌐 https://www.isse.tu-clausthal.de/ (M. Abboush)
ORCID(s): 0000-0002-5533-0029 (M. Abboush)

ability to create FDD models with an acceptable degree of interpretability.

To close this gap, this article proposes a novel approach for developing an explainable model for fault detection, identification, and localization based on hybrid DL methods and explainable artificial intelligent (XAI) techniques. The novelty of the study lies in establishing an interpretable FDD approach for real-time validation process of ASSs encompassing both single and simultaneous faults. Furthermore, a comparative study is conducted to investigate the applicability of different XAI techniques for performing model optimization. To the best of our knowledge, this is the first study to integrate XAI techniques with hybrid DL-based FDD approach, considering concurrent faults during the real-time validation phase of ASSs.

The following is a list of contributions to this work:

- An intelligent test recordings analysis approach of ASSs based on explainable hybrid deep learning methods is proposed.

- A hybrid 1dCNN-GRU-based DL model architecture is developed for fault detection, identification, and localization considering single and concurrent faults with imbalanced data.

- a white box DL model version with identification of significant variables is enabled based on interpretation of prediction results, improving the performance and efficiency along with reducing computational complexity.

- Investigating and analyzing four XAI techniques, highlighting their different computational costs and performance results.

- Employing HIL real-time simulation dataset to validate the proposed approach as an industrial application, considering user behavior and the high-fidelity ASSs models.

The structure of the article is outlined as follows: Within Section 2, a review and discussion of the relevant literature is conducted, with a focus on the similarities and differences between the present study and other related research. In Section 3, the proposed approach is presented, including the main phases and processes. The section 4 provides a comprehensive overview of the implementation structure and the case study. A detailed discussion of the results and findings can be found in Section 6. Finally, in Section 7, the study's conclusions are presented and potential future research directions are discussed.

## 2. Related Work

This section provides an overview of the most significant contributions in the field of FDD for automotive systems, with a focus on the limitations and differences between the proposed approach and related work.

The recent advancements in data-driven methodologies have yielded substantial contributions to the development of an effective FDD model, characterised by its high flexibility and scalability at minimal cost. In the automotive field, ML and DL methods have become prevalent in both industry and academia for addressing a wide range of technical challenges. The majority of the FDD models developed in the research conducted can be categorised into two phases of the ASSs life cycle, i.e., the operational phase and the development phase.

### 2.1. FDD of automotive systems during the operation phase

Due to the critical impact of faults on safety aspects during the operational phase of ASSs, i.e., autonomous vehicles, the development of a reliable FDD model has been extensively investigated by researchers. To this end, an extensive range of data-driven ML and DL algorithms have been explored for the automated detection, grouping, isolation, and identification of faults [49]. For instance, in [37], it was demonstrated how the historical data of a real vehicle could be utilised to develop an intelligent DL-based model capable of performing various diagnostic processes, including detection, isolation, identification, and prediction. Towards this end, 1dCNN and multi-class DNN-based FDD architectures were developed focusing on four types of sensor-related faults, i.e., erratic, hard-over, spike, and drift. In the same context, a novel methodology-based ensemble ML classifier was proposed in [48] to detect both known and unknown faults during analysis process of a real test drive records of the target vehicle. The robustness of the proposed model in detecting faults under different driving scenarios and fault types was demonstrated by the evaluation results, which yielded an average F2-score of 80%. At the same vehicle test level, but with a focus on simultaneous engine fault diagnosis, Wong et al. in [54] have proposed a probabilistic committee (PCM) that includes feature extraction and parameter optimization of multiple sparse Bayesian extreme learning machines (SBELMs). Assessment of the model's efficacy on actual test data set yielded an average accuracy of 86.17% in identifying both single and concurrent faults. In order to address the issue of classifying unknown faults, proposes a probabilistic fault classification algorithm that exploits the benefit of combining Weibull-calibrated single-class support vector machines and Bayesian filtering has been proposed in [24]. The proposed approach has demonstrated its superiority in diagnosing seven types of real internal combustion engine faults based on a limited training dataset with an accuracy of 85%. Within this paradigm, an advanced a data-driven DL-based diagnostic approach for short-circuit and open-circuit faults has been developed in [25] focusing on the electric vehicles (EVs) as a target system. The evaluation results on data from EV prototypes demonstrate the efficacy of the proposed LSTM architecture in comparison to alternative conventional methodologies, with an average accuracy of 97.05% and a computational time of 21.5 minutes.

Notwithstanding the considerable achievements of relevant work, the majority of the developed models exhibit deficiencies with regard to interpretability and transparency of the decisions made. To overcome the aforementioned challenges, our research, in contrast to the study cited, has examined the integration of XAI with the diagnostic approach. This approach offers a robust framework for analysing the reasons behind the prediction results and providing a superiority of the DL-based FDD model.

## 2.2. FDD of automotive systems during the development phase

During the development phase of the ASSs, it is imperative that the development process, methods and tools adhere to the requirements of ISO 26262. Consequently, there is a lack of investigation into the applicability of data-driven ML and DL methods for FDD in the development phase. Nevertheless, several studies have been conducted demonstrating the positive effect of integrating ML and DL-based FDD model into the HIL test process during the development phase. For instance, RAVEENDRAN et al. in [33] have demonstrated in their research that random forest-based FDD method offer high identification performance for vehicle brake systems. The most significant contribution of the proposed model is its capacity to detect sensor-related faults under a broad spectrum of vehicle operating conditions. In a similar manner, but for a different safety-critical automotive system, Pietrowski et al. in [31] have proposed a diagnostic algorithm for detecting incipient faults in electric power steering (EPS) under both normal and abnormal test situations. In both of the aforementioned works, HIL real-time simulations were employed to provide the training and evaluation data. With regard to the problem of detecting and identifying sensor-related faults during the real-time validation process of ASSs, novel hybrid DL-based FDD models for single and simultaneous faults were developed in [4] and [5], respectively. The innovation of the aforementioned work is predicated on the utilisation of a hybrid CNN-LSTM model for the extraction and classification of individual faults. Whereas for simultaneous faults under noise, an ensemble LSTM-RF model was developed that considers the DAE model as an initial process for noise suppression prior to classification.

Thus, it can be concluded that the mentioned studies have proposed an effective approach for developing intelligent FDD models with high accuracy and fault coverage, based on real-time HIL fault simulation. Nevertheless, the black-box nature of the models poses a challenge for optimising the model and understanding the reason behind the prediction outcome.

## 2.3. XAI-based FDD model

To address the problem of the lack of explainability of the current FDD model, several research approaches have been proposed, relying on XAI techniques [21, 50, 51]. In this regard, a range of techniques have been employed, including Integrated Gradients (IGs) [44], DeepLIFT [41], SHAP [28, 7], and LIME [53, 35].

A proportion of these focused on the use of XAI to simplify the model and reduce data dimensionality. For instance, the incorporation of the SHAP technique into CLSTM-based FDD of robot systems with a view to reducing training and inference times has been examined in [40]. The GRU-XAI FDD framework, as outlined in [13], is another example of this approach, demonstrating that the integration of an optimisation algorithm (SSA) with XAI-based feature selection can lead to a reduction in model complexity while maintaining diagnostic performance. Specifically, the study demonstrates that the employment of the SHAP technique can result in the identification of the pivotal features that significantly influence the performance of the model, achieving an impressive accuracy of 98.2%.

Other researches have investigated the applicability of XAI in explaining diagnostic results in a way that allows the causes of decision-making to be identified. In [36], for instance, the functionality of the autoencoder (AE) in detecting turbine sensor faults was extended through the use of XAI, thus enabling the analysis of the cause of the detected anomaly. The proposed ARCANA method was validated using two case studies from wind farm operators, which demonstrated the ability of XAI to provide interpretable outputs for AE. With the same objective in mind, the approach developed in [8] for the unsupervised classification of faults in rotating machines was extended to enable root cause analysis (RCA) using XAI. Specifically, SHAP and Local-DIFFI were integrated with the FDD model to identify the underlying causes of the fault, employing a strategy for ranking the importance of features. Consequently, a substantial degree of model explainability was achieved. A comparison of the results obtained from the XAI techniques employed in the three cases demonstrates that Local-DIFFI demands less computing power than SHAP.

As outlined in [43], an XAI-LCS method founded upon an Internet of Things (IoT) network was proposed with the objective of enhancing the transparency of the eXtreme Gradient Boosting model in the identification of sensor-related faults. The proposed approach facilitates the interpretation of prediction results while achieving high FDD performance with an accuracy of 99.8%. With the objective of differentiating between sensor and system faults, a hybrid CNN-XGBoost model-based fault isolation was proposed in [42]. Alongside the integration of DAE-driven fault correction, the XAI technique, i.e., SHAP, was employed to enhance the reliability of the FDD target model.

This observation indicates that the objectives of integrating XAI with FDD in the relevant work are oriented towards either optimising the developed model or performing root causes analysis (RCA). Despite the success in achieving the stated goals, previous studies have focused exclusively on individual faults to be diagnosed, ignoring the occurrence of simultaneous faults. Moreover, a paucity of research exists within the automotive sector investigating the applicability of XAI techniques in addressing the black-box nature of FDD models during operational or development phases. In order to address the existing gab in state-of-the-art FDD models of ASSs, the present study proposes a novel approach investigating a range of XAI techniques. The purpose is to enable the interpretation of the FDD of a single and simultaneous faults during real-time validation process. To this end, a hybrid CNN-GRU model was developed for fault detection, identification and localization. Furthermore, a comparative analysis was conducted on the performance of Integrated Gradients (IGs), DeepLIFT, Gradient SHAP, and DeepLIFT SHAP in terms of features selection. Real-time simulation data from the HIL system is employed in the development and testing the target model.

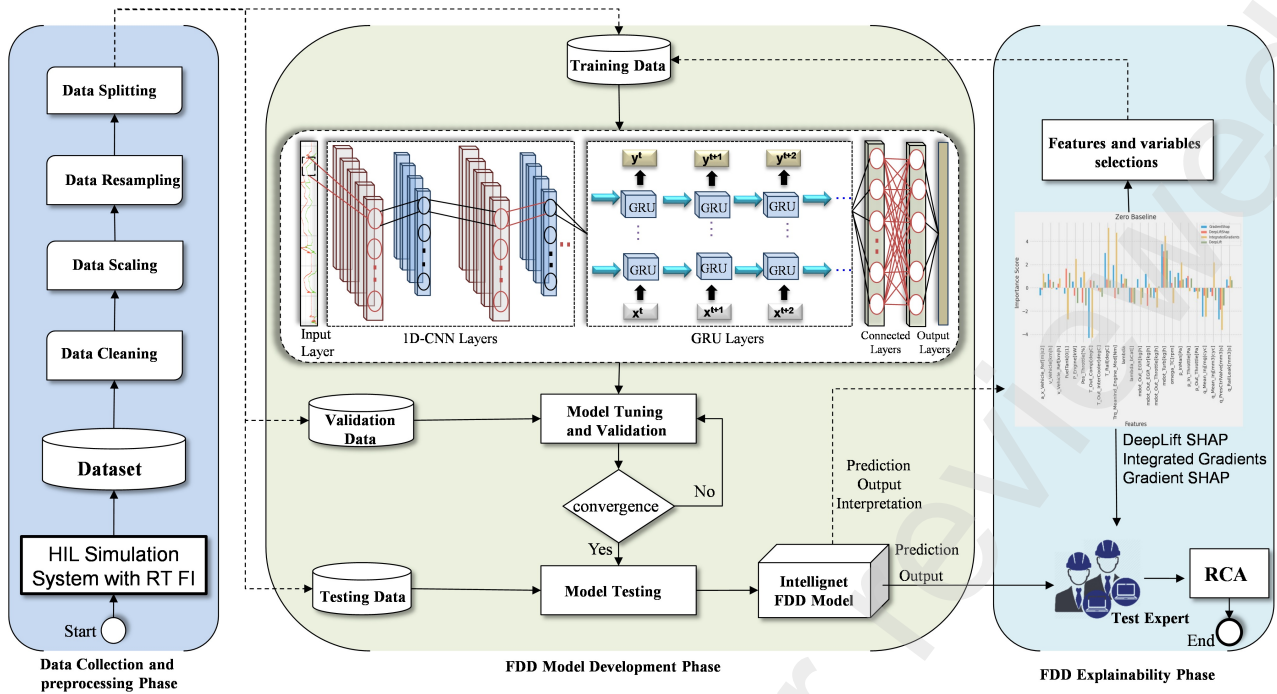

**Figure 1:** Proposed methodology for intelligent explainable hybrid DL model-based FDD approach.

## 3. Methodology

This section presents the proposed explainable intelligent FDD approach, highlighting the key processes and artefacts during model development and deployment. As illustrated in Figure 1, the proposed approach comprises three distinct phases, namely data collection and preprocessing, FDD model development, and FDD model explainability.

### 3.1. Data collection and preprocessing

Since the objective of this study is to develop an intelligent FDD model that is suitable for realistic behavior under real operating conditions, a HIL real-time simulation system was employed. Specifically, a high-fidelity vehicle system model from dSPACE is used and executed in the HIL simulator to capture the system behaviour in real-time. Furthermore, a rapid control prototype (RCP) connected to the HIL simulator via a CAN bus is utilised to execute the SUT under realistic operating conditions. It is important to note that the tester's driving behaviour was also considered when the dataset was captured using the virtual real-time test drives framework developed in [6]. Consequently, the healthy baseline dataset is captured by recording the system behaviour under fault-free conditions and taking into account the interaction between the controller and the controlled systems. Within the scope of this study, two modes have been utilised for the execution of the test drives, i.e., manual and automatic. The generation of representative fault datasets was accomplished through the utilisation of the real-time fault injection (FI) developed in [3]. The framework is centered around the idea of injecting predefined sensors and actuators-related faults into signals, accessed via the CAN bus, during real-time execution. The subsequent system response is then recorded at the system level. Consequently, realistic behaviour in the event of a fault can be recorded, with real-time constraints being taken into account. In this study, both type-related fault data and location-related fault data were collected for the development and validation of a proposed approach for two diagnostic tasks, i.e., fault identification and localization.

Subsequent to the collection and storage of the aforementioned dataset, a series of preprocessing steps are performed to enhance its quality for comprehensive analysis. As the quality of the dataset improves, the performance of the desired data-driven intelligent model is also enhanced [55]. The preprocessing steps encompass data denoising, the removal of outliers and redundant information, scaling and normalisation, resampling, and partitioning.

In real-world applications, the majority of collected data is subject to noise due to the environment of the SUT.

To improve the visibility of salient features, avoid overfitting, and ensure that the focus is on detecting the true underlying patterns rather than random fluctuations, it is therefore essential to denoise the data. In addition to the process of denoising, the elimination of irrelevant metadata and superfluous information is of paramount importance in ensuring the describability and consistency of the data. As the data set is collected based on heterogeneous components, it follows that each recorded system variable has different feature values. This, in turn, can result in poor performance or convergence issues. Therefore, a process of scaling and normalisation is applied in order to standardise the amplitudes of the system variables within a uniform range, thus ensuring that they contribute equally to the development of the model. An imbalance in the data set is a well-known challenge in model development, where the ratio of healthy and faulty samples is not equal in real-world applications [34]. In order to address this challenge, a range of resampling techniques were employed in this study, namely Random Under Sampling (RUS) [45], Class Weights (CWs) [57], and SMOTE [9]. In the context of random under sampling, the number of samples in the mature class is systematically reduced to achieve equilibrium with the samples from minority class. Unlike RUS, CWs serve to augment the effect of the minority class by being weighted more heavily without altering the number of samples. Consequently, the model's focus is oriented towards the samples with high weighting, while the valuable data of the majority class is not disregarded. Finally, the SMOTE technique is employed to generate synthetic samples in the minority class by interpolation, thus facilitating the training process from a more diverse set of examples. In order to facilitate the model's capacity to discern temporal patterns and dependencies within the data, a data segmentation process is implemented. The implementation of this process provides fixed-size sequences for the purpose of efficient training, which are referred to as "windows". The final step in the preprocessing stage is the division of the data into three distinct components: training, validation, and testing. A central tenet of this approach is the objective of avoiding overfitting, whilst ensuring that the model is appropriately verified with unseen data.

### 3.2. FDD model development phase

Following the preprocessing of the data set, the subsequent phase is that of FDD model development. The development phase includes the training process, optimisation and validation, and model testing.

During the development process, the selection of the model architecture and the definition of the hyperparameters are of the utmost importance. The selection of DL methods is typically determined by the characteristics of the dataset and the technical problem to be solved, such as classification, clustering, or regression. Within the context of this study, the task under investigation is delineated as a multi-class, multi-label classification problem of time series data [52]. Recurrent neural networks (RNNs), LSTM, Gated recurrent unit (GRU), and 1dCNN have demonstrated their efficacy in processing substantial amounts of sequential data with temporal dependencies for classification tasks [20]. However, RNN technique is susceptible to the vanishing gradient problem, computationally intensive, and lack mechanisms for storing information [29]. LSTM technique is characterised by a high level of complexity stemming from their numerous gates, a propensity to overfit, and a substantial demand for computational costs in terms of time and resources [19]. 1dCNN has been shown to be effective at feature extraction; however, it lacks the ability to explicitly model temporal dependencies, and it exhibits poor detection patterns when using fixed-size filters. Despite its computational efficiency and comparatively simple architecture in comparison to alternative techniques, GRU encounters difficulties with very long sequences [46]. Consequently, a hybrid 1dCNN-GRU-based model architecture is proposed that leverages the complementary strengths and overcomes the limitations of each technique. Specifically, the 1dCNN convolution operation is employed to efficiently extract representative features and capture local patterns, while GRUs are utilised to effectively capture temporal dependencies with minimal computational expense. In order to facilitate the mapping between the learned features and the desired number of output classes, an fully connected (FC) layer is incorporated after the GRU layers. The architecture of the 1dCNN-GRU for fault type classification model (FTCM) and fault localization model (FLM) are illustrated in Tables 1 and 2, respectively.

Subsequent to the input layer, five convolutional layers are constructed, including Conv1d, BatchNorm1d, ReLU, and MaxPool1d. These layers utilise filters to systematically search the data, thereby enabling the model to discover localized patterns such as trends, irregularities, and recurring cycles. This mechanism is of particular significance in the identification of latent structures within data samples that may not be evident in their original state. The aforementioned layers are able to transform the data into a more abstract and meaningful representation by capturing these localised patterns. The mathematical formulation of the convolution process is represeted as follows:

$$\mathbf{X}_{\text{out}} = \sigma(\mathbf{W}_c * \mathbf{X}_{\text{in}} + \mathbf{b}_c)$$

where $\mathbf{X}_{\text{in}}$ is the input time-series, $\mathbf{W}_c$ represents the convolution filter weights, $*$ denotes the convolution operation, $\mathbf{b}_c$ is the bias term, $\sigma$ is the activation function (e.g., ReLU), and $\mathbf{X}_{\text{out}}$ is the output feature map.

Subsequent to the CNN layers, the development of two GRU layers is undertaken in order to effectively capture the temporal dependencies in the dataset. The fundamental principle of GRU is predicated on the utilisation of gating mechanisms, thereby facilitating the regulated flow of information over time. It is worth noting that GRU depends on detailed, representative features that are extracted sequentially by the CNN layer. This, in turn, has been shown to improve the learning process of temporal dynamics and long-term

**Table 1**
Layer-wise Summary of the FTCM Model Architecture

| Layer (Type) | Output Shape | Parameters (#) |
|---|---|---|
| Conv1d-1 | [-1, 32, 500] | 2,336 |
| BatchNorm1d-2 | [-1, 32, 500] | 64 |
| ReLU-3 | [-1, 32, 500] | 0 |
| MaxPool1d-4 | [-1, 32, 499] | 0 |
| Conv1d-5 | [-1, 64, 499] | 6,208 |
| BatchNorm1d-6 | [-1, 64, 499] | 128 |
| ReLU-7 | [-1, 64, 499] | 0 |
| MaxPool1d-8 | [-1, 64, 498] | 0 |
| Conv1d-9 | [-1, 128, 498] | 24,704 |
| BatchNorm1d-10 | [-1, 128, 498] | 256 |
| ReLU-11 | [-1, 128, 498] | 0 |
| MaxPool1d-12 | [-1, 128, 497] | 0 |
| Conv1d-13 | [-1, 256, 497] | 98,560 |
| BatchNorm1d-14 | [-1, 256, 497] | 512 |
| ReLU-15 | [-1, 256, 497] | 0 |
| MaxPool1d-16 | [-1, 256, 248] | 0 |
| GRU-17 | [[-1, 248, 512], [-1, 2, 512]] | 0 |
| Linear-18 | [-1, 128] | 65,664 |
| ReLU-19 | [-1, 128] | 0 |
| Dropout-20 | [-1, 128] | 0 |
| Linear-21 | [-1, 7] | 903 |

*Total Parameters: 199,335    |    Trainable: 199,335    |    Non-Trainable: 0*

dependencies, thereby enabling effective tracking of time-dependent patterns. The GRU gate can be mathematically represented by the following equations:

$$z_t = \sigma \left( \mathbf{W}_z \mathbf{h}_{t-1} + \mathbf{U}_z \mathbf{X}_t + \mathbf{b}_z \right)$$

$$r_t = \sigma \left( \mathbf{W}_r \mathbf{h}_{t-1} + \mathbf{U}_r \mathbf{X}_t + \mathbf{b}_r \right)$$

$$\tilde{\mathbf{h}}_t = \tanh \left( \mathbf{W}_h (r_t \cdot \mathbf{h}_{t-1}) + \mathbf{U}_h \mathbf{X}_t + \mathbf{b}_h \right)$$

$$\mathbf{h}_t = (1 - z_t) \cdot \mathbf{h}_{t-1} + z_t \cdot \tilde{\mathbf{h}}_t$$

where $\mathbf{X}_t$ is the input at time $t$, $\mathbf{h}_{t-1}$ is the previous hidden state, $\mathbf{W}_z, \mathbf{W}_r, \mathbf{W}_h$ are weight matrices for the update, reset, and candidate states, $\mathbf{U}_z, \mathbf{U}_r, \mathbf{U}_h$ are input-related weight matrices, $\mathbf{b}_z, \mathbf{b}_r, \mathbf{b}_h$ are biases for the gates and candidate state, and $\sigma$, tanh are activation functions for the gating and candidate computations.

In conclusion, the proposed model identifies the FC layer as the final component for combining the representations learned by the GRU into a suitable format for an accurate decision-making and prediction process. Mathematically, the FC layer is represented by the following equation:

$$\mathbf{y} = \sigma \left( \mathbf{W}_f \mathbf{h}_{\text{GRU}} + \mathbf{b}_f \right)$$

where $\mathbf{h}_{\text{GRU}}$ is the output from the final GRU layer, $\mathbf{W}_f$ is the weight matrix of the fully connected (FC) layer that maps the GRU outputs to the target dimension, $\mathbf{b}_f$ is the bias term of the FC layer, $\sigma$ is the activation function (e.g., softmax for multi-class or sigmoid for binary), and $\mathbf{y}$ is the predicted output, representing class probabilities.

For multi-class classification, the softmax activation is applied:

$$\mathbf{y}_i = \frac{e^{\mathbf{z}_i}}{\sum_j e^{\mathbf{z}_j}}$$

where $\mathbf{z}_i$ is the logit (raw score) for class $i$ and $\mathbf{y}_i$ is the predicted probability for class $i$.

Upon completion of the preliminary model architecture, the initial and range values of the hyperparameters are determined. Depending on the range determined, the hyperparameters are adjusted, leading to the optimisation of the model structure. To this end, the model's performance is evaluated based on its hyperparameters in terms of accuracy, precision, recall, and F1-score, using validation data. In contradistinction to the utilisation of manual random tuning, this study employs automatic tuning mechanisms to systematically explore the space of hyperparameters. Grid search, random search, and bayesian optimization are used to systematically evaluate model performance under different configurations and to perform tuning accordingly, taking into account the trade-off between model accuracy and computational resources. Convergence is achieved through the selection of optimal hyperparameters, thus completing the validation process. The optimally validated model is then verified using unseen test data to ensure that the model

**Table 2**
Layer-wise Summary of the FLM Architecture

| Layer (Type) | Output Shape | Parameters (#) |
|---|---|---|
| Conv1d-1 | [-1, 32, 500] | 2,336 |
| BatchNorm1d-2 | [-1, 32, 500] | 64 |
| ReLU-3 | [-1, 32, 500] | 0 |
| MaxPool1d-4 | [-1, 32, 499] | 0 |
| Conv1d-5 | [-1, 64, 499] | 6,208 |
| BatchNorm1d-6 | [-1, 64, 499] | 128 |
| ReLU-7 | [-1, 64, 499] | 0 |
| MaxPool1d-8 | [-1, 64, 498] | 0 |
| Conv1d-9 | [-1, 128, 498] | 24,704 |
| BatchNorm1d-10 | [-1, 128, 498] | 256 |
| ReLU-11 | [-1, 128, 498] | 0 |
| MaxPool1d-12 | [-1, 128, 497] | 0 |
| Conv1d-13 | [-1, 256, 497] | 98,560 |
| BatchNorm1d-14 | [-1, 256, 497] | 512 |
| ReLU-15 | [-1, 256, 497] | 0 |
| MaxPool1d-16 | [-1, 256, 496] | 0 |
| Conv1d-17 | [-1, 512, 496] | 393,728 |
| BatchNorm1d-18 | [-1, 512, 496] | 1,024 |
| ReLU-19 | [-1, 512, 496] | 0 |
| MaxPool1d-20 | [-1, 512, 495] | 0 |
| GRU-21 | [[-1, 495, 512], [-1, 2, 512]] | 0 |
| Linear-22 | [-1, 128] | 65,664 |
| ReLU-23 | [-1, 128] | 0 |
| Dropout-24 | [-1, 128] | 0 |
| Linear-25 | [-1, 7] | 903 |

*Total Parameters: 594,087    |    Trainable: 594,087    |    Non-Trainable: 0*

meets the requirements in terms of generalisation without overfitting.

### 3.3. FDD model explainability

In the third phase of the proposed approach, the interpretability attribute is activated and added to the trained FDD model from the previous phase. In order to achieve this objective, four XAI techniques are employed, i.e., Integrated Gradients (IGs), Deep Learning Important Features (DeepLIFT), Gradient SHapley Additive exPlanations (Gradient SHAP), and DeepLIFT SHAP.

The primary objective of IG [44] is to calculate assignments by tracking the gradients from a designated starting point, such as a zero vector, to the specific input. This process involves the evaluation of the change in output as each input feature gradually shifts from its initial value to its actual value. Mathematically, The attribution of each feature is calculated by integrating gradients along the path from the baseline $x'$ to the input $x$:

$$IG_j(x) = (x_j - x'_j) \times \int_{\alpha=0}^{1} \frac{\partial f(x' + \alpha \cdot (x - x'))}{\partial x_j} \, d\alpha$$

Where:

- $f(x)$ is the model's prediction for input $x$,
- $x'$ is the baseline input,
- $x_j$ and $x'_j$ are the $j$-th features of $x$ and $x'$,
- $\alpha$ interpolates between $x'$ and $x$,
- $\frac{\partial f}{\partial x_j}$ is the gradient with respect to the $j$-th feature.

The final attribution vector is:

$$IG(x) = (x - x') \odot \int_{\alpha=0}^{1} \frac{\partial f(x' + \alpha \cdot (x - x'))}{\partial x} \, d\alpha$$

Where $\odot$ denotes the element-wise product.

DeepLIFT [41] employs a layer-wise method for assigning relevance in order to evaluate the change in the model's output relative to a specific reference point. In particular, the contribution of each neuron in the network is decomposed and traced back to the input features. Mathematically, the attributions for each feature $x_j$ are computed by comparing activations at the reference and actual input. For layer $l$, the difference in activation $\Delta A_{l,j}$ is given by:

$$\Delta A_{l,j} = A_{l,j}(x) - A_{l,j}(x')$$

Where:

- $A_{l,j}(x)$ and $A_{l,j}(x')$ are the activations of the $j$-th unit in layer $l$ for the input $x$ and reference $x'$, respectively.

- $\Delta A_{l,j}$ is the activation difference.

Attributions are then computed as:

$$\text{Attribution}_j(x) = \sum_l \left(\text{Relevance Score}_{l,j} \times \Delta A_{l,j}\right)$$

Where Relevance Score$_{l,j}$ reflects the layer's contribution to the final output, depending on network architecture and activation relationships.

Gradient SHAP [28] is a method that approximates Shapley values by employing gradients to determine assignments and evaluate the importance of features. This approach facilitates efficient and effective computation, while also accounting for the uncertainty of predictions. The mathematical equation of Gradient SHAP is represented as the following equation:

$$\hat{v}_j = \mathbb{E}_{S \sim \pi}\left[\left(f(x_S \cup j) - f(x_S)\right) \nabla_{x_j} f(x_S \cup j)\right]$$

Where:

- $f(x)$ is the model's prediction for input $x$,

- $x_S$ is the input with a subset of features excluding the $j$-th feature,

- $x_S \cup j$ includes the $j$-th feature added to $x_S$,

- $\nabla_{x_j} f(x_S \cup j)$ is the gradient of the model's prediction with respect to the $j$-th feature.

In order to leverage the advantages of DeepLIFT and SHAP, DeepLIFT efficiently approximates SHAP values in order to address the challenge of interactions and nonlinear relationships in the model. In contrast to the standalone DeepLIFT approach, the DeepLIFT SHAP [28] technique considers numerous baselines, thereby facilitating enhanced capture of a diverse array of interactions and feature importance. Consequently, it is regarded as the optimal solution to the trade-off between understanding model behaviour and computational efficiency, particularly in the context of real-world applications. DeepLIFT SHAP provides a detailed analysis of the influence that individual features have on the prediction process for different fault classes within the target system. The identification of the most significant features associated with each type of fault or its location enables tests to be focused on the most important attributes, thus increasing test efficiency and effectiveness. From a model perspective, DeepLIFT SHAP improves interpretability by revealing the reasons behind predictions, promoting trust, and supporting system diagnosis. Mathematically, the equations of is represented as the following:

$$\hat{v}_j = \mathbb{E}_{S \sim \pi}\left[\left(f(x_S \cup j) - f(x_S)\right) \cdot \Delta_{x_j} f(x_S \cup j)\right]$$

Where:

- $f(x)$ is the model's prediction for input $x$,

- $x_S$ is a subset of input features excluding feature $j$,

- $x_S \cup j$ is the input with feature $j$ added to the subset $x_S$,

- $\Delta_{x_j} f(x_S \cup j)$ is the DeepLIFT contribution for feature $j$, which represents the difference in activations for input $x_S \cup j$ relative to the baseline.

The attribution for feature $j$ is computed as:

$$\hat{v}_j = \sum_{S \subseteq F \setminus j} \frac{|S|!(|F| - |S| - 1)!}{|F|!} \left(f(x_S \cup j) - f(x_S)\right) \cdot \Delta_{x_j} f(x_S \cup j)$$

Where $F$ is the full set of features and $S$ is a subset excluding $j$.

The aforementioned techniques are utilised to evaluate the contribution of features at the sample level. This approach supports informed decision-making, reduces overfitting, and improves model performance.

For deeper understanding of model behavior, the XAI techniques are determined by three attributes, including Global Feature Importance (GFI), Per-Class Feature Importance (PCFI), and Feature Interactions (FIs).

The core idea behind GFI is to identify the features that have the greatest impact on the prediction process, thereby ensuring a high level of model performance and reliability. In our case, GFI is mainly used to determine specific features that contribute to faults. In other words, the trained model can be optimized by focusing on the important features that lead to optimal prediction performance. This approach ensures that only the most critical features are given due consideration during the development process, thereby minimising complexity and ensuring comprehensive coverage and robustness. Conversely, features that are redundant or irrelevant can be disregarded, thereby circumventing the necessity for additional data analysis.

PCFI provides detailed analyses of how individual features influence the prediction process for different fault classes within the target system. By identifying the most important features associated with each fault type or its location, tests can be focused on the most effective attributes, increasing test efficiency and effectiveness. From a model perspective, PCFI improves interpretability by revealing the reasons behind the prediction, promoting trust, and supporting system diagnosis. Unlike traditional feature weighting methods, PCFI can be adapted for both local and global interpretations and is therefore suitable for multi-class classification problems. Thus, PCFI contributes to more accurate fault identification, increases diagnostic effectiveness, and improves model clarity and refinement.

Finally, feature interactions (FIs) play a pivotal role in comprehending feature interactions when the effect of a feature on the target variable cannot be fully understood without considering the other interacting features. Notably, this method is distinguished by its capacity to reveal non-linear dependencies and joint effects of features that are pivotal for determining the system status under faulty conditions. with a view to determining the nature of these interactions, the influence of each feature is first evaluated in order to ascertain its individual contribution. By simultaneously adjusting feature pairs and evaluating their joint contribution, FIs then identifies interaction effects. A substantial increase in the joint contribution in comparison to the sum of the separate contributions signifies a robust interaction and emphasises the underlying dependencies that exert influence on the model's outcomes. By comprehending these intricate characteristics, the interactions and nuanced relationships that determine the model predictions can be identified. Consequently, FIs are regarded as the optimal solution for scenarios where faults occur concurrently.

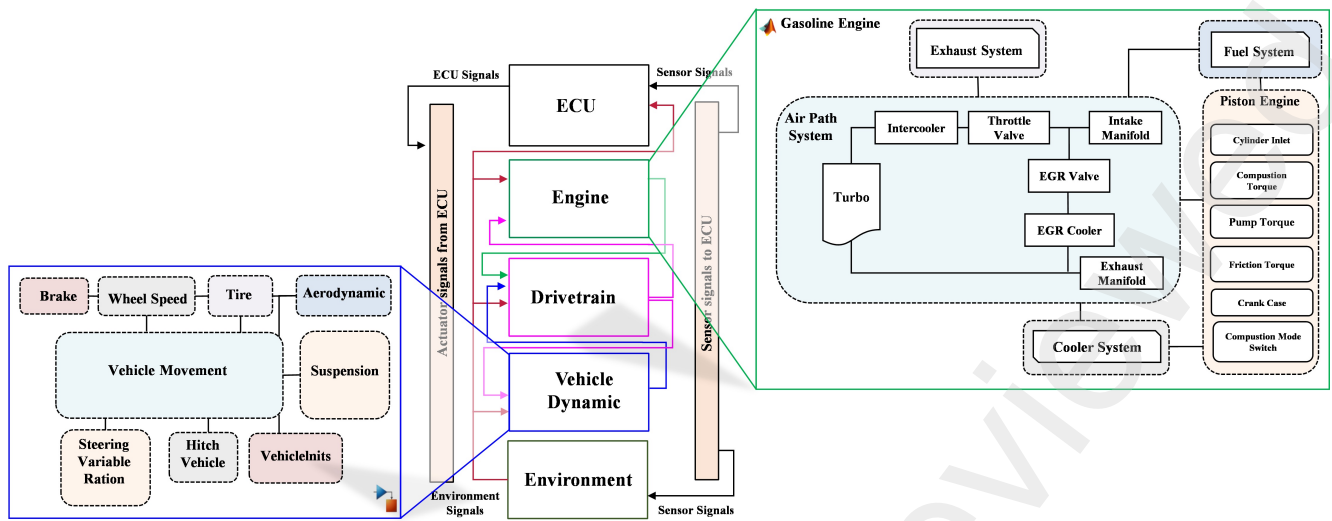

**Figure 2:** Automotive simulation model architecture of the case study.

## 4. Case study and Experimental Implementation

### 4.1. ASM Gasoline Engine and HIL simulation

The research explores the function of a gasoline engine using dSPACE's Automotive Simulation Models (ASM) [15], a toolkit for simulating and analyzing automotive systems. ASM enables engineers to improve performance, efficiency, and control systems through detailed simulations. It provides an in-depth model of a turbocharged 6-cylinder engine, supporting both direct and manifold injection systems, as well as manual and automatic transmissions. The simulation accurately models key engine components, including the fuel system, air path, cooling system, and exhaust system. The high-fildelity ASM gasoline engine has been developed and verified in MATLAB-Simulink environment [2]. The architecture of the system model used in this study is presented in Figure 2

At the fundamental hardware level, a dSPACE HIL configuration is employed, comprising the HIL SCALEXIO simulator [14] and MicroAutoBox II [16], interconnected via a CAN bus system. The MicroAutoBox II serves as a RCP simulating a control unit by executing the necessary control logic.In this study, the RCP is considered a SUT and employs a DS1401 base board, and contains a 900 MHz processor, a 6th generation Intel® Core™ i7-6822EQ chip, and 16 MB of memory, to execute the control logic in real-time. On the other hand, SCALEXIO facilitates the real-time simulation of the entire vehicle model, thereby offering a distinct advantage in terms of functionality and operational efficiency. The communication of data signals between the simulator and the RCP is facilitated by the CAN bus protocol, a system that encompasses both sensor inputs and control outputs. The HIL system facilitates precise real-time simulation of complex systems, enabling embedded control units to undergo rigorous testing under realistic conditions. Moreover, it mitigates the risks associated with hardware testing, a particularly valuable benefit in scenarios where physical testing can be costly, time-consuming, or hazardous. The HIL environment incorporates steering wheel and pedals, enabling engineers to execute manual test drives in accordance with prescribed test requirements. The elements of the virtual test drives based on HIL simulations are illustrated in Figure 3.

## 5. Data preprocessing

Two datasets are created based on three driving scenarios: highway driving, lane changes, and city driving. The first dataset focuses on fault types, capturing various fault categories such as noise (F1), gain (F2), and offsets (F3), as well as their combinations. The difference between healthy and faulty data in the engine speed and engine torque signal in case of single fault occurrence is illustrated in 4 and 5, respectively. On the other hand, the effect of concurrent faults on the engine speed and engine speed is illustrated in 6 and 7, respectively. The second dataset concentrates on identifying fault locations within the system, including the acceleration position pedal sensor (L1), steering wheel angle sensor (L2), and engine speed sensor (L3), along with their combinations. Each dataset contains 24 features. Table 3 provides an overview of the key features collected during this process. The metadata is removed from the generated files, and the feature values are standardized. After standardizing the data, random undersampling and SMOTE are used to address class imbalance. Table 4 shows the sample counts for the original, undersampled, and SMOTE-oversampled data. The data is then windowed into segments by sliding across the data with a fixed step size. Then the data is splitted into three subsets: 70% for training, 15% for validation, and 15 % for testing.

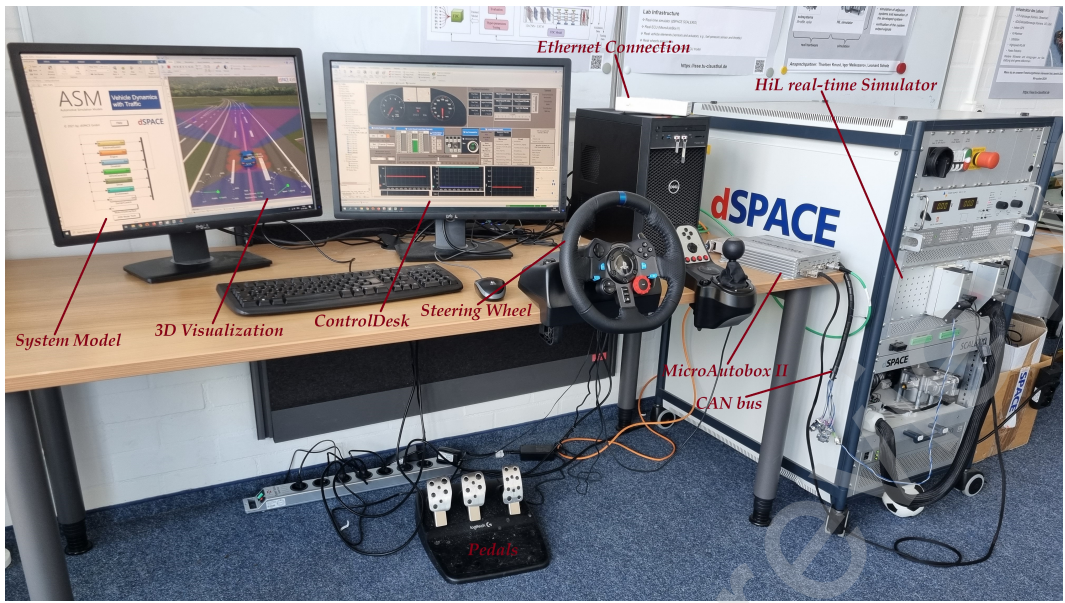

**Figure 3:** HIL-based virtual test drives for real-time validation.

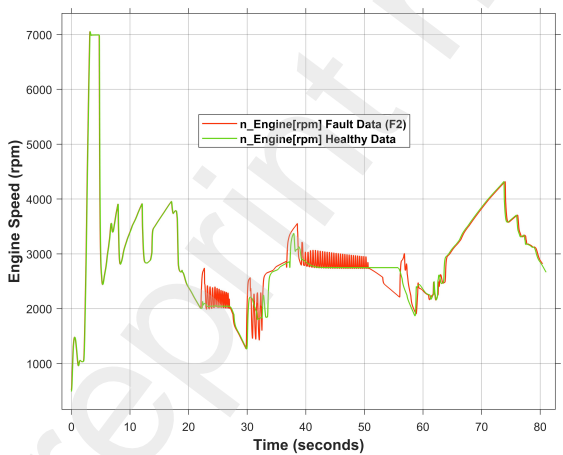

**Figure 4:** Engine speed under single fault.

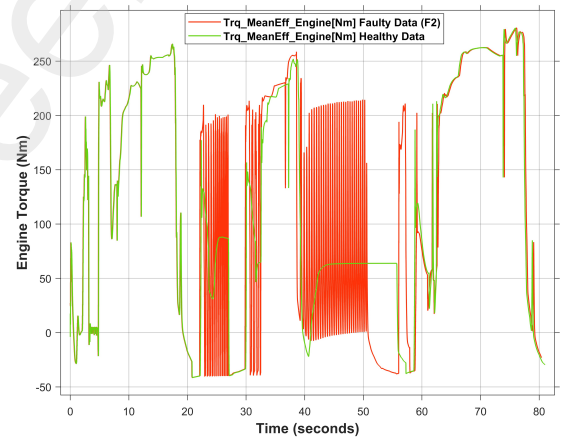

**Figure 5:** Engine torque under single fault.

## 5.1. Model Development

The proposed 1dCNN-GRU model consists of a sequential architecture that integrates CNNs, GRUs, and FC layers. The model begins with CNN layers, which extract spatial features from the input data by applying convolutional filters to capture local patterns. Following the CNN layers, GRU layers are used to address the temporal aspects of the data, enabling the model to learn and maintain long-term dependencies.

This sequential configuration allows the model to effectively combine spatial and temporal information. The architecture concludes with FC layers, which integrate the features learned from both the CNN and GRU components, allowing for a unified prediction. Additionally, simple RNN, LSTM, and GRU models are employed as benchmarks to

skip=10pt

**Table 3**
Overview of the Collected Dataset: Feature Names, and Descriptions

| Feature Name | Description |
|---|---|
| a_x_Vehicle_Ref[m/s$^2$] | Target vehicle acceleration. |
| v_Vehicle[km/h] | Current vehicle speed. |
| v_Vehicle_Ref[km/h] | Predefined speed for comparison. |
| FuelTank[0,1] | Fuel remaining (0: empty, 1: full). |
| P_Engine[kW] | Output power of the engine. |
| Pos_Throttle[%] | Position of the throttle valve. |
| T_Out_Comp[°C] | Compressor outlet temperature. |
| T_Out_InterCooler[°C] | Intercooler outlet temperature. |
| T_Rail[°C] | Temperature in the fuel rail. |
| Trq_MeanInd_Engine_Mod[Nm] | Average indicated torque. |
| lambda | Air-fuel ratio. |
| lambda_bCat[] | Lambda at the catalytic converter. |
| mdot_Out_EGR[kg/h] | Mass flow rate of exhaust gas. |
| mdot_Out_EGR_Air[kg/h] | Air mass flow from EGR. |
| mdot_Out_Throttle[kg/h] | Air mass flow at throttle valve. |
| mdot_Turb[kg/h] | Mass flow rate exiting turbine. |
| omega_TC[rpm] | Turbocharger speed (rpm). |
| p_InMan[Pa] | Pressure in the intake manifold. |
| p_In_Throttle[Pa] | Pressure at throttle inlet. |
| p_Out_Throttle[Pa] | Pressure at throttle outlet. |
| q_Mean_Inj[mg/cycle] | Average fuel injection quantity. |
| q_Mean_Inj_Alt[mm$^3$/cycle] | Alternate fuel injection quantity. |
| q_PresCtrlValve[mm$^3$/s] | Flow through pressure control valve. |
| q_RailLeak[mm$^3$/s] | Leakage flow in the fuel rail. |

**Table 4**
Data Distribution Across Fault Locations and Types: Original, Undersampled, and SMOTE

| | Fault Location Data Count | | | | Fault Type Data Count | | |
|---|---|---|---|---|---|---|---|
| Location | O | U | SM | Type | O | U | SM |
| H | 232,899 | | | H | 231,698 | | |
| L1 | 298,397 | | | F1 | 240,899 | | |
| L2 | 244,901 | | | F2 | 245,796 | | |
| L3 | 237,001 | 232,899 | 298,397 | F3 | 239,198 | 231,698 | 256,998 |
| L1L2 | 254,700 | | | F1F2 | 243,399 | | |
| L1L3 | 246,298 | | | F1F3 | 256,998 | | |
| L2L3 | 252,996 | | | F2F3 | 249,299 | | |

systematically evaluate and compare the performance of different model architectures.

To optimize the performance of the proposed model, a systematic hyperparameter tuning approach is employed. The hyperparameter space includes window size, step size, number of CNN layers, number of GRU layers, hidden layer sizes, number of FC layers, and resampling techniques. This tuning is conducted for both the fault location and fault type models.

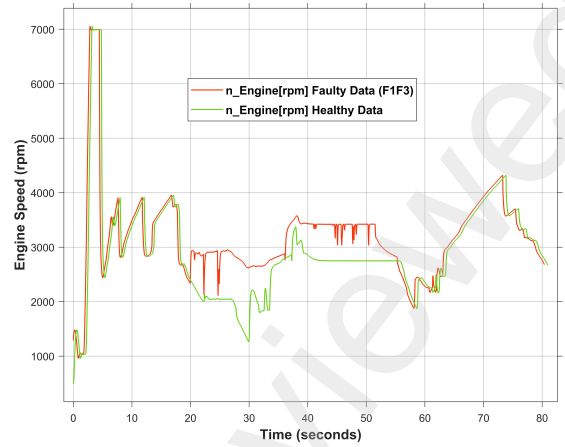

**Figure 6:** Engine speed under concurrent faults.

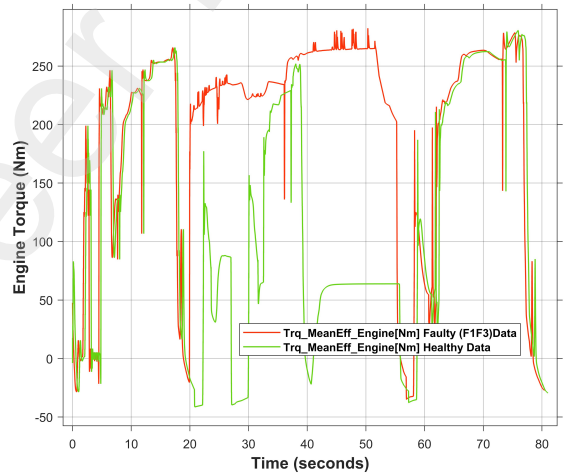

**Figure 7:** Engine torque under concurrent faults.

The architecture of the proposed models, i.e., FLM and FTCM, begin with a sequence of Conv1D layers to extract spatial features, with the number of channels increasing progressively (FLM: 32 to 256; FTCM: 24 to 512). While both use small kernel sizes, the FTCM incorporates a stride of 2 in deeper layers to enhance feature abstraction. Each model integrates a GRU module to capture temporal dependencies—three layers for FLM and two for FTCM—each with 512 hidden units. Finally, both models employ fully connected layers to reduce the feature space to 128 and then to 7, yielding class probabilities for the final output.

## 5.2. Hyperparameters Tuning

In deep learning, optimizing performance relies heavily on effective model training and careful hyperparameter tuning. Table 5 summarizes the key hyperparameters used across different trials, providing a consistent foundation for comparison.

**Table 5**
Shared Hyperparameters and Their Values Across All Trials

| Hyperparameter | Value |
|---|---|
| Learning Rate | 0.001 |
| Batch Size | 1024 |
| Number of Epochs | 256 |
| Optimizer | Adam |
| Dropout Rate | 0.3 |
| L1 | 0.0001 |
| L2 | 0.0001 |
| Patience | 8 |
| Scheduler | Cosine Annealing Scheduler |
| Sampling Technique | None |

Table 6 outlines the key hyperparameters for the neural network components, including CNN, GRU, and FC layers, along with training parameters. It lists a range of values for each hyperparameter, such as layer counts, hidden sizes, resampling techniques, and window/step sizes, which will be tested to assess the model's performance and identify the optimal configurations.

### 5.3. XAI Integration

XAI methods such as IG, DeepLIFT, Gradient SHAP, and DeepLIFT SHAP are employed to compute GFI 4, PCFI 10, and FIs 17. GFI offers an overall measure of how individual features influence the model's predictions across the entire dataset by aggregating feature attributions computed for each sample. PCFI builds on this by highlighting the contribution of features specific to each class, aiding in understanding class-wise decision behavior. FIs aim to reveal the combined effect of two or more features on the model's output, which exceed the sum of their individual impacts, offering deeper insight into feature interactions.

---

**Algorithm 1 GFI Computation**

**Input:** Trained model $M$, dataset $D$, baseline type $B$ (*zero*, *mean*, *median*, or *random*)

**Output:** Global Feature Importance (GFI) scores

1 **Procedure** ComputeGlobalImportance()**:**

2      Ensure $M$ supports gradient-based attribution  Select baseline ( $B \in$ {Zero, Mean, Median, Random} **for** *each sample* $x \in D$ **do**

3          Compute attributions using IG, DeepLIFT, Gradient SHAP, or DeepLIFT SHAP

4      Aggregate attributions across $D$ (mean) to obtain GFI  Compare results across methods and baselines for validation

---

**Algorithm 2 PCFI Computation**

**Input:** Trained model $M$, labeled dataset $D$, baseline $B$ (e.g., zero, mean, domain-specific)

**Output:** Per-Class Feature Importance (PCFI) scores

5 **Procedure** ComputeClassWiseImportance()**:**

6      Ensure $M$ supports class-specific gradient-based attribution  Select baseline ( $B \in$ {zero, mean, domain-specific} **for** *each sample* $x \in D$ **do**

7          Compute attributions via IG, DeepLIFT, Gradient SHAP, or DeepLIFT SHAP

8      **for** *each class* $c \in C$ **do**

9          Group samples in $D$ with label $c$  Compute mean attributions for each feature in class $c$

10      Compare per-class scores to identify shared/unique features

---

**Algorithm 3 FIs Computation**

**Input:** Model $M$, dataset $D$

**Output:** Feature Interaction (FIs) scores

11 **Procedure** ComputeFeatureInteractions()**:**

12      Ensure $M$ supports gradient-based attribution methods  **for** *each feature* $f_i \in D$ **do**

13          Compute attribution using IG, DeepLIFT, Gradient SHAP, or DeepLIFT SHAP

14      **for** *each feature pair* $f_i, f_j \subset D$ **do**

15          Perturb $f_i$ and $f_j$ jointly, compute combined attribution  **if** *combined score > sum of individual scores* **then**

16              Mark $(f_i, f_j)$ as interacting

17      Visualize interactions (e.g., via heatmap)

---

## 6. Result and Discussion

### 6.1. Evaluation Metrics

To evaluate the proposed model and compare it with benchmarks, several metrics are used, i.e., precision, recall, F1-score and accuracy [32]. The confusion matrix shows true and false classifications. Accuracy measures overall correctness, precision indicates the proportion of correct positive predictions, and recall reflects how well actual positives are identified. The F1-score balances precision and recall.

### 6.2. Model Evaluations

Figure 8 shows the model's performance over 250 epochs. The loss history plot indicates steady decreases in both training and validation losses, stabilizing around 0.2, which suggests effective learning and minimal overfitting. The accuracy history plot shows a consistent increase in both training and validation accuracies, reaching about 0.9 by the end, with the curves closely aligned, indicating stable and continuous improvement throughout the training process.

Figure 9 shows the model's classification performance across seven classes, with high accuracy and effective discrimination. The normalized confusion matrix reveals strong performance, with L1L3 achieving perfect accuracy, while L2 has the lowest at 0.88. The probability histogram indicates high confidence in predictions, and the precision-recall curve shows perfect average precision for most classes. The ROC curve displays an AUC of 1.00 for all classes, demonstrating excellent sensitivity and specificity.

**Table 6**
Key Hyperparameters for Neural Network Architecture Components

| Component | Hyperparameter | Range | FLM Optimal Values | FTCM Optimal Values |
|---|---|---|---|---|
| Convolutional Layers | Number of Conv Layers | 1 to 7 | 4 | 5 |
| GRU Layers | Number of GRU Layers | 1 to 4 | 3 | 2 |
| | Hidden Size | 32, 64, 256, 512 | 512 | 512 |
| FC Layers | Number of FC Layers | 1 to 7 | 1 | 1 |
| Training Parameters | Resampling Technique | None, Undersample, SMOTE | SMOTE | Undersample |
| | Window Size | 10, 50, 100, 500, 1000 | 500 | 50 |
| | Step Size | 10, 50, 100, 500, 1000 | 10 | 100 |

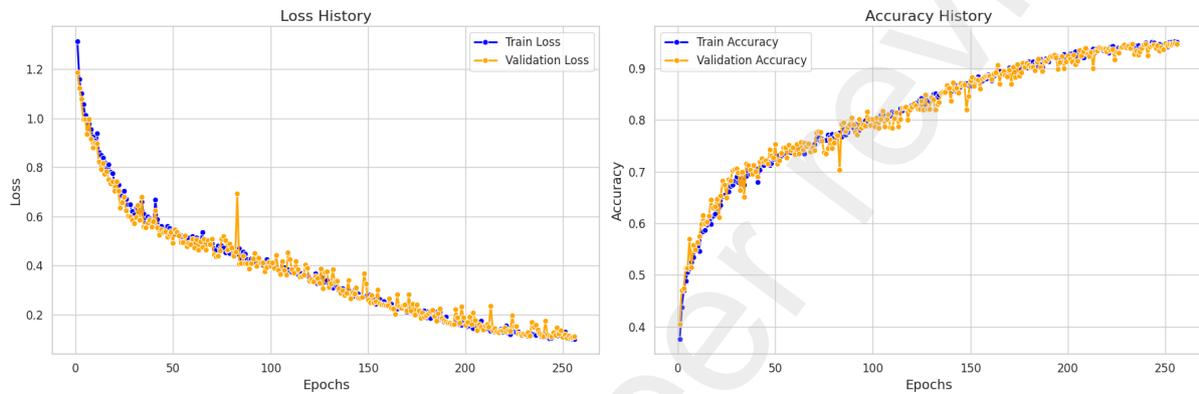

**Figure 8:** Training and Validation Loss and Accuracy for the FLM over 250 Epochs

Table 7 provides a comparative analysis of four deep learning architectures—RNN, LSTM, GRU, and the proposed FLM—evaluated using standard classification metrics and training time. The RNN model demonstrated the weakest performance, achieving an accuracy of only 43.07%, with similarly low precision (43.10%), recall (43.06%), and F1-score (40.94%). These results highlight RNN's limited capacity to model long-term dependencies due to its vanishing gradient issues. LSTM improved significantly, with 57.73% accuracy, leveraging its gating mechanisms to better capture temporal dynamics. Its precision (61.30%), recall (57.70%), and F1-score (57.87%) indicate more reliable performance across classes. GRU offered further gains, reaching 74.45% accuracy and well-balanced precision (74.84%), recall (74.37%), and F1-score (74.45%), reflecting its efficiency in handling sequence data with fewer parameters than LSTM. The proposed FLM model achieved state-of-the-art results, scoring 97.40% across all metrics, demonstrating strong representational capacity and robustness in sequence modeling. However, this came at the cost of a substantially higher training time (22,998.75 seconds), compared to RNN (373.92s), LSTM (699.93s), and GRU (1356.55s). These findings underscore the trade-off between model accuracy and computational complexity.

Figure 10 shows the model's classification performance across seven classes (H, F1, F2, F3, F1F2, F1F3, F2F3), with high accuracy and effective discrimination. The normalized confusion matrix highlights strong performance, with most diagonal values close to 1.00, although some misclassifications in F3 indicate minor overlap. The probability histogram demonstrates high confidence in predictions, with most probabilities near 0 or 1, while intermediate distributions reflect lower certainty. The precision-recall curve reveals near-perfect average precision scores, indicating strong precision and recall with minimal false positives. The ROC curve shows nearly perfect scores for all classes, underscoring the model's robust ability to differentiate between true and false positives.

Table 8 compares the classification performance and computational efficiency of the same baseline models alongside the proposed FTCM. RNN achieved moderate accuracy (69.25 %) but exhibited the longest training time among the standard models, making it less efficient for practical deployment. LSTM, while slightly lower in accuracy (67.99%), offered a more balanced precision and recall profile, indicating improved class-wise consistency. GRU significantly outperformed both, achieving 85.66% accuracy along with superior precision, recall, and F1-scores, demonstrating its effectiveness in sequence modeling tasks. The proposed FTCM delivered the best overall performance, attaining 97.19% accuracy and nearly identical values across all evaluation metrics, which suggests highly stable and reliable predictions. However, this performance came with the highest training time (7896.56 seconds), reflecting the increased computational demands of the model. This comparison highlights the critical balance between predictive accuracy and training efficiency when selecting deep learning architectures for real-world applications.

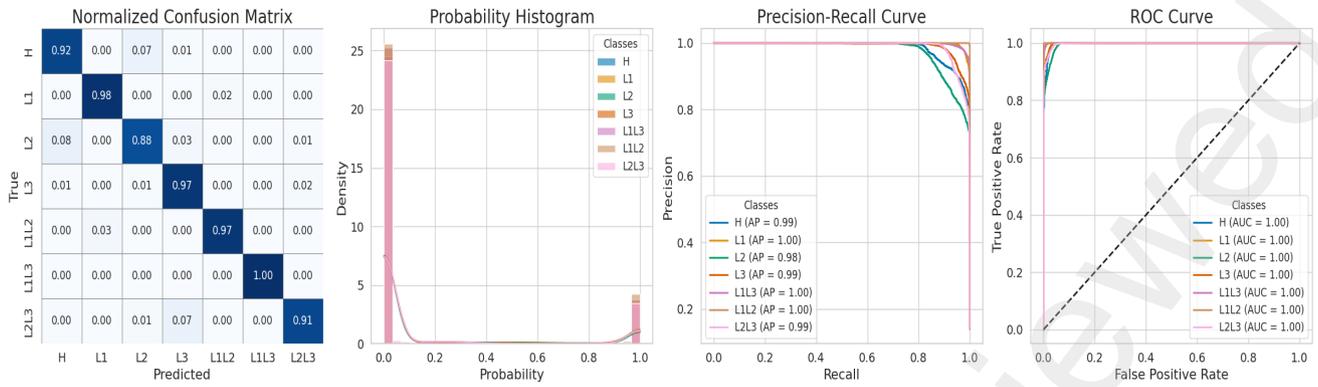

**Figure 9:** Evaluation Plots Showing FLM's Classification Performance

**Table 7**
Performance Metrics for the Baseline Models (RNN, LSTM, GRU) and FLM

| Metric | RNN | LSTM | GRU | FLM |
|---|---|---|---|---|
| Accuracy | 43.07 | 57.73 | 74.45 | **97.40** |
| Precision | 43.10 | 61.30 | 74.84 | **97.40** |
| Recall | 43.06 | 57.70 | 74.37 | **97.40** |
| F1-Score | 40.94 | 57.87 | 74.45 | **97.40** |
| Train Time (s) | **373.92** | 699.93 | 1356.55 | 22998.75 |
| Test Time (s) | 7.36 | 7.77 | 7.92 | **5.01** |

### 6.3. XAI

*Global Feature Importance:* the GFI results for FLM, comparing attribution methods (`IGs`, `DeepLIFT`, `Gradient SHAP`, `DeepLIFT SHAP`) and baseline methods (`zero`, `mean`, `median`, `random`) are presented in Figure 11 . `IGs` displayed moderate and stable importance scores, focusing on steady feature influence, especially under the `zero` baseline. `DeepLIFT` showed sharper, higher-magnitude scores, reflecting its sensitivity to baseline changes and making it responsive to features with strong contrasts. `Gradient SHAP` generated a wider distribution of scores, emphasizing subtle feature relationships and variations in data, while DeepLIFT SHAP provided more balanced and consistent scores. The baseline comparisons revealed the `zero` baseline emphasizing linear relationships (e.g., P_Engine[kW], T_Out_Comp[°C]), while the `random` baseline led to broader variability, highlighting features like a_x_Vehicle_Ref[m/s²] and lambda_bCat[].

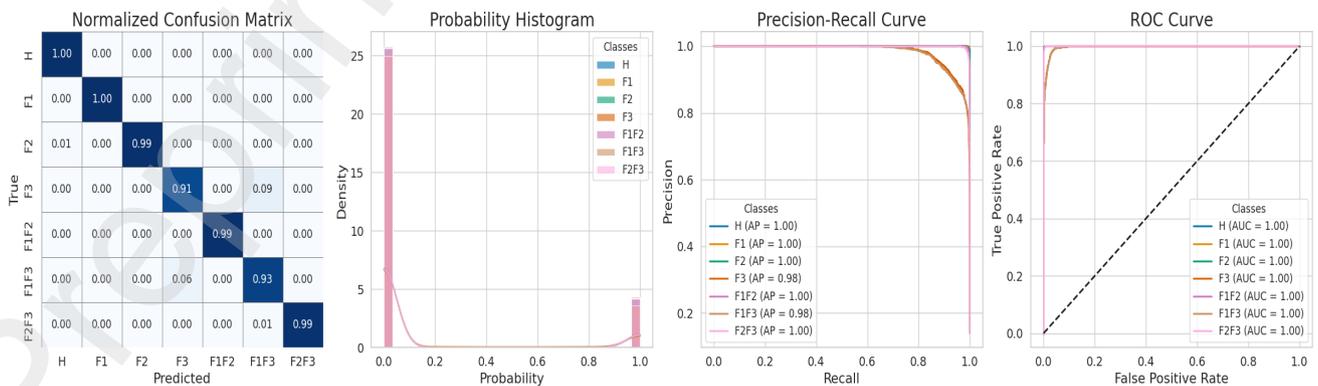

**Figure 10:** Evaluation Plots Showing FTCM's Classification Performance

**Table 8**
Performance Metrics for Baseline Models (RNN, LSTM, GRU) and FTCM

| Metric | RNN | LSTM | GRU | FTCM |
|---|---|---|---|---|
| Accuracy | 69.25 | 67.99 | 85.66 | **97.19** |
| Precision | 67.96 | 68.64 | 86.98 | **97.22** |
| Recall | 69.10 | 67.77 | 85.57 | **97.21** |
| F1-Score | 66.38 | 66.17 | 85.76 | **97.21** |
| Train Time (s) | 3885.03 | **2493.00** | 4702.98 | 7896.56 |
| Test Time (s) | 7.08 | 6.65 | **5.84** | 6.10 |

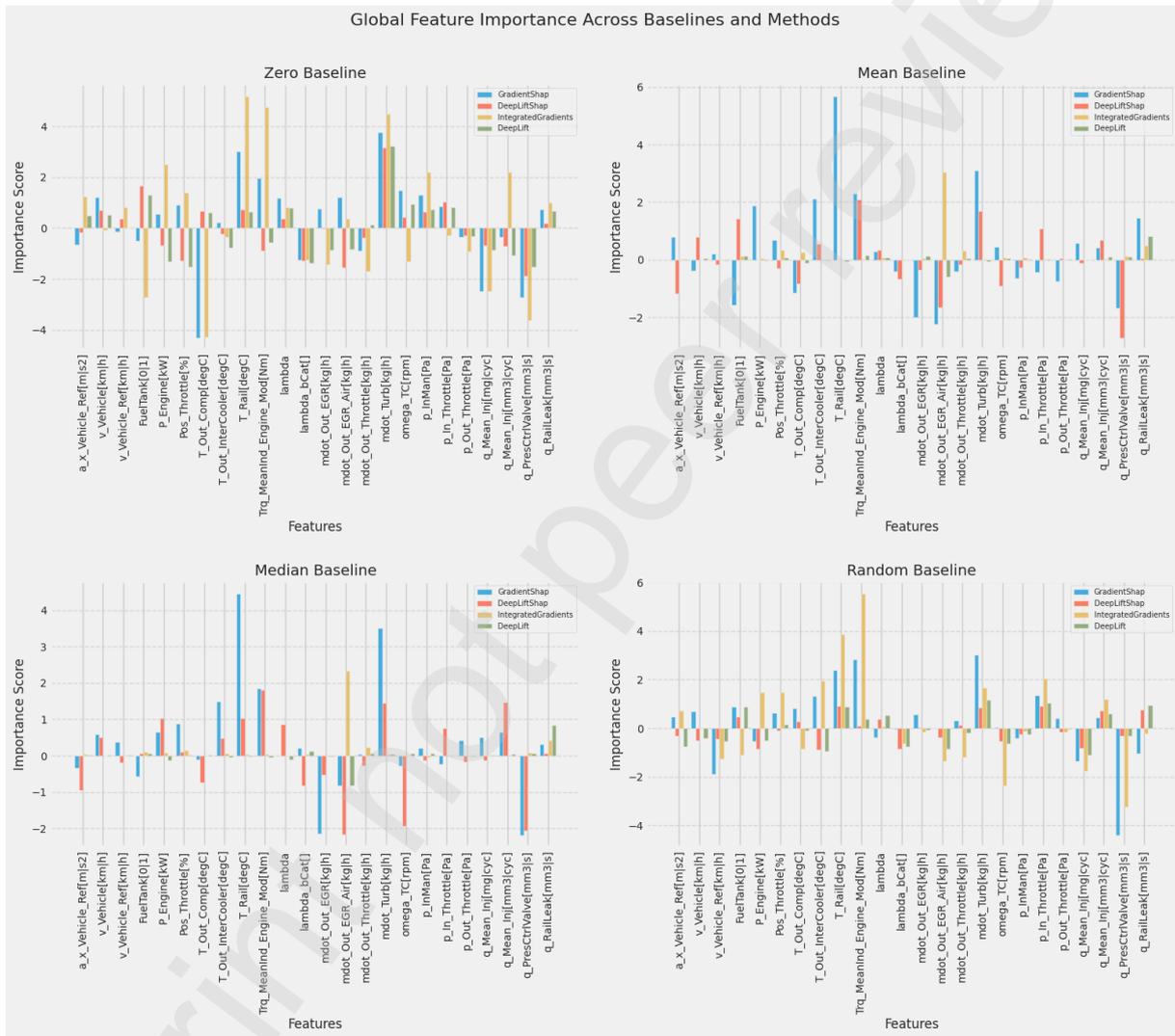

**Figure 11:** GFI of FLM Interpretability Methods and Across Baselines

Table 9 ranks the important features influencing FLM predictions based on GFI metrics.

After retraining the FLM model using these features, performance in Table 10 showed a slight reduction ( 2%) in accuracy, precision, recall, and F1-score. However, the retrained model significantly reduced training time from 22,999s to 5,413s and showed a slight reduction in test time, improving computational efficiency.

*Per-Class Feature Importance* the PCFI results for the FLM, comparing attribution methods (IGs, DeepLIFT, Gradient

**Table 9**
Ranked Features for the FLM via IGs, DeepLIFT, Gradient SHAP, and DeepLIFT SHAP

| Rank | Feature |
|------|---------|
| 1 | P_Engine[kW] |
| 2 | q_PresCtrlValve[mm3\|s] |
| 3 | lambda |
| 4 | T_Out_Comp[degC] |
| 5 | q_Mean_Inj[mg\|cyc] |
| 6 | v_Vehicle[km\|h] |
| 7 | a_x_Vehicle_Ref[m\|s2] |
| 8 | T_Out_InterCooler[degC] |
| 9 | Pos_Throttle[%] |
| 10 | mdot_Out_Throttle[kg\|h] |

skip=10pt

**Table 10**
Performance Comparison of Original and Retrained FLM

| Model | Accuracy (%) | Precision (%) | Recall (%) | F1-Score (%) | Train Time (s) | Test Time (s) |
|-------|--------------|---------------|------------|--------------|----------------|---------------|
| **Original FLM** | **97.40** | **97.40** | **97.40** | **97.40** | 22998.75 | 5.01 |
| **Retrained FLM** | 95.62 | 95.60 | 95.60 | 95.60 | **5412.72** | **4.54** |

SHAP, DeepLIFT SHAP) across various baselines (zero, mean, median, random) are presented in Figure 6.3.

IGs exhibit low variability, making them suitable for conservative feature importance estimation, though they underemphasize subtle contributions. DeepLIFT offers more stability than Gradient SHAP but is influenced by baseline choice, with the zero baseline often inflating feature importance. Gradient SHAP highlights extreme feature contributions with high variability, making it useful for detecting sharp contrasts but prone to overestimation. DeepLIFT SHAP combines the strengths of DeepLIFT and SHAP, offering smoother, more interpretable results with consistent feature rankings across baselines.

For baseline comparisons, the zero baseline tends to exaggerate feature importance, particularly for Gradient SHAP and DeepLIFT. The mean baseline smooths distributions, with DeepLIFT SHAP emphasizing features like mdot_Out_Throttle[kg/h]. The median baseline provides balanced results, highlighting features like q_Mean_Inj[mm³/cyc]. The random baseline introduces high variability, especially for features like P_Engine[kW], with Gradient SHAP capturing these fluctuations more than other methods.

In per-class comparisons, class H shows high variability, with DeepLIFT SHAP identifying Pos_Throttle[%] and p_InMan[Pa] as key features. For classes L1, L2, and L3, DeepLIFT SHAP provides smoother trends, highlighting features like mdot_Turb[kg/h]. In combined classes (L1L2, L1L3, L2L3), DeepLIFT SHAP and IGs show consistent importance for features like q_RailLeak[mm³/s] and lambda_bCat[].

*FIs:* the Feature Interactions (FIs) matrix using the Gradient SHAP attribution method with the mean baseline is illustrated in Figure 13, highlighting key interactions between features in the model's predictions.

Strong positive interactions include a high correlation (0.91) between lambda_bCat[] and mdot_Out_EGR[kg/h], both of which significantly influence emissions. omega_TC[rpm] (turbo speed) shows perfect self-correlation (1.0) and a strong relationship with mdot_Turb[kg/h] (0.90), as well as a moderate correlation with p_Out_Throttle[Pa] (0.44). Additionally, FuelTank[0|1] and Pos_Throttle[%] exhibit a moderate interaction (0.52), reflecting fuel management during acceleration.

Negative interactions include an inverse relationship (-0.54) between lambda and mdot_Out_EGR_Air[kg/h], indicating opposing effects in air-fuel dynamics. q_RailLeak[mm³/s] shows moderate negative interactions with mdot_Out_Throttle [kg|h] (-0.42) and omega_TC[rpm] (-0.49), suggesting a negative relationship with throttle airflow and turbo speed.

Neutral interactions are observed between q_Mean_Inj [m³/cyc], q_Mean_Inj[mg/cyc], and most other features, as well as weak interactions for a_x_Vehicle_Ref[m|s²] (vehicle acceleration), indicating minimal influence on or from other features.

*Trade-off:* the interpretability versus complexity trade-off is a key consideration in model development, balancing the simplicity and ease of understanding offered by less complex models with the high performance delivered by more sophisticated, but harder-to-interpret, models.

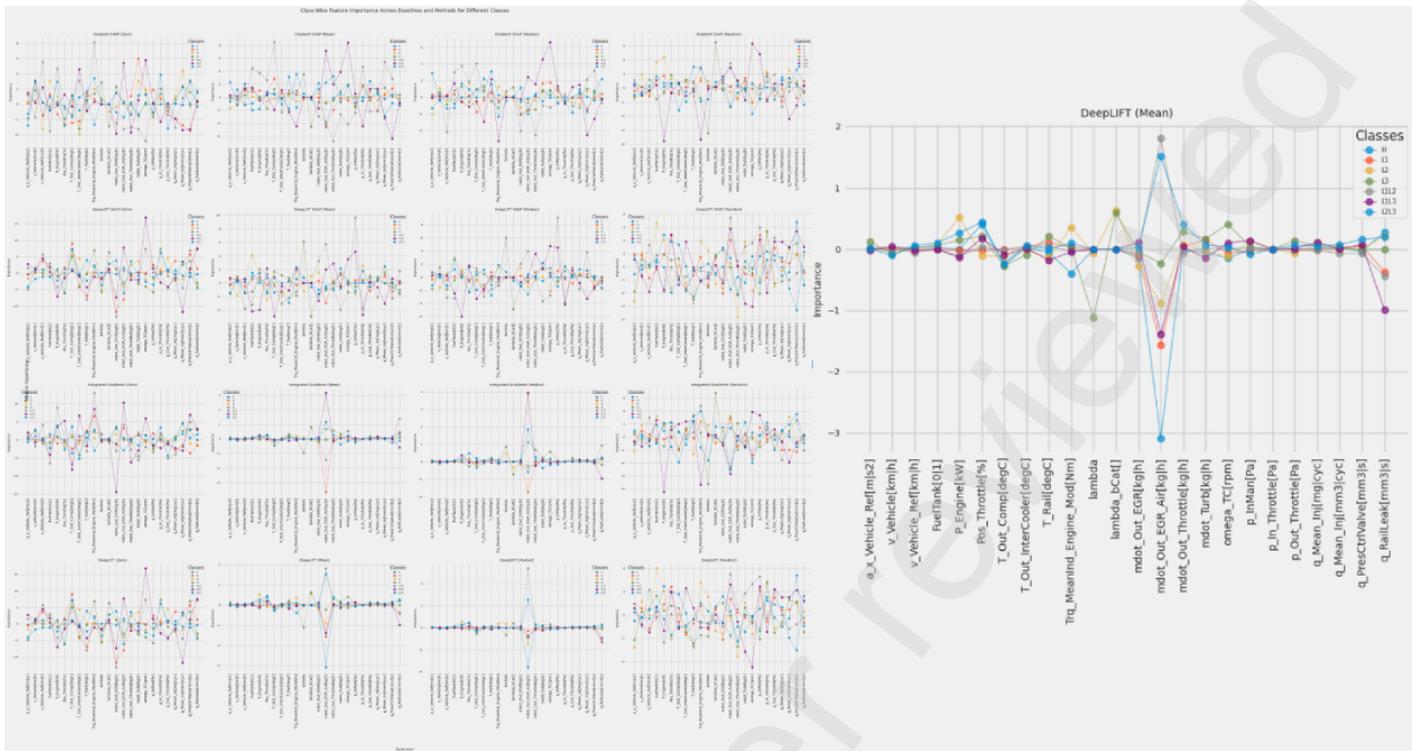

**Figure 12:** PCFI of FLM Interpretability Methods and Baselines

skip=10pt

**Table 11**
Comparison of Attribution Methods Across Fault Location Models (Times in seconds)

| Model | IGs | DeepLIFT | Gradient SHAP | DeepLIFT SHAP |
|---|---|---|---|---|
| **RNN** | 10.42 | **0.92** | **1.75** | 12.92 |
| **LSTM** | 8.71 | 3.28 | 3.76 | 11.27 |
| **GRU** | **8.24** | 3.53 | 3.77 | **10.23** |
| **Original FLM** | 19.56 | 3.65 | 3.53 | 32.83 |
| **Retrained FLM** | 19.29 | 3.55 | 3.65 | 32.19 |

Attribution method execution times for fault localization models are compared in Table 11, which illustrates the time (in seconds) taken by four attribution methods across five models. DeepLIFT is the fastest method in most cases, particularly for the RNN model, which takes only 0.92 seconds. IGs and Gradient SHAP are slower, especially for the RNN (10.42 seconds) and GRU (8.24 seconds). DeepLIFT SHAP consistently takes the longest time across all models, with the original FLM and retrained FLM models taking approximately 32 seconds each.

## 7. Conclusion and Future Work

To address the challenge of developing a reliable, efficient, and interpretable FDD model for analysing test records during the real-time validation process of ASSs, a novel methodology was proposed in this article. In particular, a data-driven hybrid 1dCNN-GRU model architecture was developed that leverages the advantages of both methods in representative feature extraction and the capture of temporal dependencies in sequential data. Furthermore, XAI techniques were integrated into the proposed approach, enabling the provision of a white-box FDD model capable of determining the reason behind the prediction results. To this end, four XAI techniques, i.e, IGs, DeepLIFT, Gradient SHAP, and DeepLIFT SHAP, were compared and investigated. To verify the effectiveness of the proposed approach, a real-time simulation dataset from a HIL-based virtual test drive was utilised. This dataset is based on highly realistic automotive system models and takes user behaviour into account.

The evaluation results demonstrate the superiority of the proposed model over state-of-the-art FDD models for both fault type identification and fault localization tasks, with F1-scores of 97.40% and 97.19%, respectively. Furthermore, the proposed hybrid 1dCNN-GRU model exhibited a superior performance in terms of accuracy for single and concurrent sensor-related faults when evaluated on unseen data, with an average accuracy of 96%.

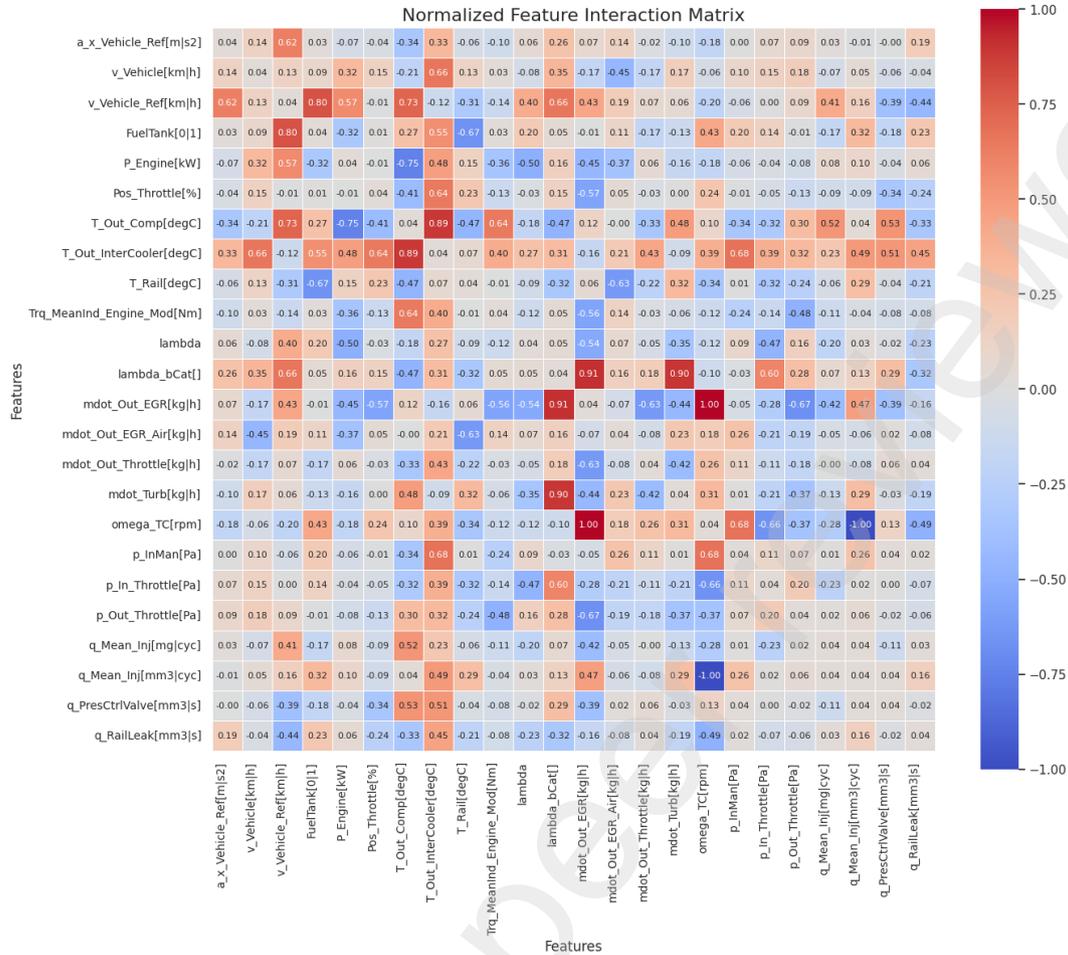

**Figure 13:** Normalized FIs Heatmap

The experimental results demonstrate that the implementation of XAI facilitates a reduction in model complexity while ensuring performance by adjusting the model based on feature weighting. Specifically, based on the results of the XAI baselines (i.e., zero, mean, median and random), the number of features is reduced from 24 to 10. Consequently, the computing costs are optimized, with reduced training and testing time. Furthermore, the XAI results offer a valuable instrument to assist engineers in conducting RCA by quantifying the significant contribution of features to the occurrence of signal and concurrent faults. The findings of the comparison of XAI methods reveal that DeepLIFT and Gradient SHAP are the most efficient methods with low computational costs, while IGs SHAP and DeepLIFT SHAP require high processing time. It is therefore concluded that the proposed approach has the capacity to overcome the limitations of the current black-box FDD model and improve the prediction process with high explainability. This, in turn, not only contributes to improving the validation process of ASSs and supporting safety engineers, but also ensures optimized computing costs for real-time applications.

In future research, the proposed approach can be expanded to investigate other applications of XAI techniques, such as trust estimation, human-AI collaboration, and identification of model biases. Moreover, the issue of automatically executing RCA of concurrent faults can be further examined based on the preliminary outcomes of our approach.

## References

[1] . . Iso 26262-10:2018-road vehicles—functional safety—part 10: Guideline on iso 26262. https://www.iso.org/standard/68392.html. Accessed on: Feb. 14, 2025, [Online].

[2] , . Simulink - simulation und model-based design (o. j.): in: Matlab simulink. https://de.mathworks.com/products/simulink.html. (Accessed on 02/22/2025).

[3] Abboush, M., Bamal, D., Knieke, C., Rausch, A., 2022a. Hardware-in-the-loop-based real-time fault injection framework for dynamic behavior analysis of automotive software systems. Sensors 22, 1360.

[4] Abboush, M., Bamal, D., Knieke, C., Rausch, A., 2022b. Intelligent fault detection and classification based on hybrid deep learning methods for hardware-in-the-loop test of automotive software systems. Sensors 22, 4066.

[5] Abboush, M., Knieke, C., Rausch, A., 2023. Intelligent identification of simultaneous faults of automotive software systems under noisy and imbalanced data using ensemble lstm and random forest. IEEE Access 11, 140022–140040.

[6] Abboush, M., Knieke, C., Rausch, A., 2024. A virtual testing framework for real-time validation of automotive software systems based on hardware in the loop and fault injection. Sensors 24, 3733.

[7] Ahmed, S., Kaiser, M.S., Hossain, M.S., Andersson, K., 2024. A comparative analysis of lime and shap interpreters with explainable ml-based diabetes predictions. IEEE Access .

[8] Brito, L.C., Susto, G.A., Brito, J.N., Duarte, M.A., 2022. An explainable artificial intelligence approach for unsupervised fault detection and diagnosis in rotating machinery. Mechanical Systems and Signal Processing 163, 108105.

[9] Chawla, N.V., Bowyer, K.W., Hall, L.O., Kegelmeyer, W.P., 2002. Smote: synthetic minority over-sampling technique. Journal of artificial intelligence research 16, 321–357.

[10] Chen, J., Zhang, L., Li, Y., Shi, Y., Gao, X., Hu, Y., 2022. A review of computing-based automated fault detection and diagnosis of heating, ventilation and air conditioning systems. Renewable and Sustainable Energy Reviews 161, 112395.

[11] Chi, C.F., Sigmund, D., Astardi, M.O., 2020. Classification scheme for root cause and failure modes and effects analysis (fmea) of passenger vehicle recalls. Reliability Engineering & System Safety 200, 106929.

[12] Chi, Y., Dong, Y., Wang, Z.J., Yu, F.R., Leung, V.C., 2022. Knowledge-based fault diagnosis in industrial internet of things: a survey. IEEE Internet of Things Journal 9, 12886–12900.

[13] Dave, V., Borade, H., Agrawal, H., Purohit, A., Padia, N., Vakharia, V., 2024. Deep learning-enhanced small-sample bearing fault analysis using q-transform and hog image features in a gru-xai framework. Machines 12, 373.

[14] dSPACE, 2023. SCALEXIO System. URL: https://www.dspace.com/en/pub/home/products/hw/simulator_hardware/scalexio.cfm.

[15] dSPACE, 2025a. ASM Gasoline Engine. URL: https://www.dspace.com/de/gmb/home/products/sw/automotive_simulation_models/produkte_asm/asm_engine_models/asm_gasoline_engine_simulation.cfm.

[16] dSPACE, 2025b. MicroAutoBox II Embedded PC. URL: https://www.dspace.com/en/pub/home/products/hw/micautob/microautobox_embedded_pc.cfm.

[17] Gao, Z., Cecati, C., Ding, S.X., 2015. A survey of fault diagnosis and fault-tolerant techniques—part i: Fault diagnosis with model-based and signal-based approaches. IEEE transactions on industrial electronics 62, 3757–3767.

[18] Garousi, V., Felderer, M., Karapıçak, Ç.M., Yılmaz, U., 2018. Testing embedded software: A survey of the literature. Information and Software Technology 104, 14–45.

[19] Gers, F.A., Schraudolph, N.N., Schmidhuber, J., 2002. Learning precise timing with lstm recurrent networks. Journal of machine learning research 3, 115–143.

[20] Han, Z., Zhao, J., Leung, H., Ma, K.F., Wang, W., 2019. A review of deep learning models for time series prediction. IEEE Sensors Journal 21, 7833–7848.

[21] Harinarayan, R.R.A., Shalinie, S.M., 2022. Xfddc: explainable fault detection diagnosis and correction framework for chemical process systems. Process Safety and Environmental Protection 165, 463–474.

[22] Isermann, R., 2005. Model-based fault-detection and diagnosis–status and applications. Annual Reviews in control 29, 71–85.

[23] Jordan, C.V., Hauer, F., Foth, P., Pretschner, A., 2020. Time-series-based clustering for failure analysis in hardware-in-the-loop setups: An automotive case study, in: 2020 IEEE International Symposium on Software Reliability Engineering Workshops (ISSREW), IEEE. pp. 67–72.

[24] Jung, D., 2020. Data-driven open-set fault classification of residual data using bayesian filtering. IEEE Transactions on Control Systems Technology 28, 2045–2052.

[25] Kaplan, H., Tehrani, K., Jamshidi, M., 2021. A fault diagnosis design based on deep learning approach for electric vehicle applications. Energies 14, 6599.

[26] Karl, I., Berg, G., Ruger, F., Farber, B., 2013. Driving behavior and simulator sickness while driving the vehicle in the loop: Validation of longitudinal driving behavior. IEEE intelligent transportation systems magazine 5, 42–57.

[27] Klein, S., Savelsberg, R., Xia, F., Guse, D., Andert, J., Blochwitz, T., Bellanger, C., Walter, S., Beringer, S., Jochheim, J., et al., 2017. Engine in the loop: Closed loop test bench control with real-time simulation. SAE International journal of commercial vehicles 10, 95–105.

[28] Lundberg, S.M., Lee, S.I., 2017. A unified approach to interpreting model predictions. Advances in neural information processing systems 30.

[29] Mienye, I.D., Swart, T.G., Obaido, G., 2024. Recurrent neural networks: A comprehensive review of architectures, variants, and applications. Information 15, 517.

[30] von Neumann-Cosel, K., Dupuis, M., Weiss, C., 2009. Virtual test drive-provision of a consistent tool-set for [d, h, s, v]-in-the-loop, in: Proceedings of the driving simulation conference Monaco.

[31] Pietrowski, W., Puskarczyk, M., Szymenderski, J., 2024. Fault detection methods for electric power steering system using hardware in the loop simulation. Energies 17, 3486.

[32] Powers, D.M., 2020. Evaluation: from precision, recall and f-measure to roc, informedness, markedness and correlation. arXiv preprint arXiv:2010.16061 .

[33] Raveendran, R., Devika, K., Subramanian, S.C., 2020. Brake fault identification and fault-tolerant directional stability control of heavy road vehicles. IEEE Access 8, 169229–169246.

[34] Ren, Z., Lin, T., Feng, K., Zhu, Y., Liu, Z., Yan, K., 2023. A systematic review on imbalanced learning methods in intelligent fault diagnosis. IEEE Transactions on Instrumentation and Measurement 72, 1–35.

[35] Ribeiro, M.T., Singh, S., Guestrin, C., 2016. " why should i trust you?" explaining the predictions of any classifier, in: Proceedings of the 22nd ACM SIGKDD international conference on knowledge discovery and data mining, pp. 1135–1144.

[36] Roelofs, C.M., Lutz, M.A., Faulstich, S., Vogt, S., 2021. Autoencoder-based anomaly root cause analysis for wind turbines. Energy and AI 4, 100065.

[37] Safavi, S., Safavi, M.A., Hamid, H., Fallah, S., 2021. Multi-sensor fault detection, identification, isolation and health forecasting for autonomous vehicles. Sensors 21, 2547.

[38] Sahu, A.R., Palei, S.K., Mishra, A., 2024. Data-driven fault diagnosis approaches for industrial equipment: A review. Expert Systems 41, e13360.

[39] Shafiq, S., Mashkoor, A., Mayr-Dorn, C., Egyed, A., 2021. A literature review of using machine learning in software development life cycle stages. IEee Access 9, 140896–140920.

[40] Shojaeinasab, A., Jalayer, M., Baniasadi, A., Najjaran, H., 2024. Unveiling the black box: A unified xai framework for signal-based deep learning models. Machines 12, 121.

[41] Shrikumar, A., Greenside, P., Kundaje, A., 2017. Learning important features through propagating activation differences, in: International conference on machine learning, PMIR. pp. 3145–3153.

[42] Sinha, A., Das, D., 2023a. An explainable deep learning approach for detection and isolation of sensor and machine faults in predictive maintenance paradigm. Measurement Science and Technology 35, 015122.

[43] Sinha, A., Das, D., 2023b. Xai-lcs: Explainable ai-based fault diagnosis of low-cost sensors. IEEE Sensors Letters 7, 1–4.

[44] Sundararajan, M., Taly, A., Yan, Q., 2017. Axiomatic attribution for deep networks, in: International conference on machine learning, PMLR. pp. 3319–3328.

[45] Tahir, M.A., Kittler, J., Yan, F., 2012. Inverse random under-sampling for class imbalance problem and its application to multi-label classification. Pattern Recognition 45, 3738–3750.

[46] Tao, Q., Liu, F., Li, Y., Sidorov, D., 2019. Air pollution forecasting using a deep learning model based on 1d convnets and bidirectional gru. IEEE access 7, 76690–76698.

[47] Theissler, A., 2014. Anomaly detection in recordings from in-vehicle networks. Big data and applications 23, 26.

[48] Theissler, A., 2017. Detecting known and unknown faults in automotive systems using ensemble-based anomaly detection. Knowledge-Based Systems 123, 163–173.

[49] Theissler, A., Pérez-Velázquez, J., Kettelgerdes, M., Elger, G., 2021. Predictive maintenance enabled by machine learning: Use cases and challenges in the automotive industry. Reliability engineering & system safety 215, 107864.

[50] Theissler, A., Spinnato, F., Schlegel, U., Guidotti, R., 2022. Explainable ai for time series classification: a review, taxonomy and research directions. Ieee Access 10, 100700–100724.

[51] Thomas, A.W., Ré, C., Poldrack, R.A., 2023. Benchmarking explanation methods for mental state decoding with deep learning models. Neuroimage 273, 120109.

[52] Tsoumakas, G., Katakis, I., 2008. Multi-label classification: An overview. Data Warehousing and Mining: Concepts, Methodologies, Tools, and Applications , 64–74.

[53] Visani, G., Bagli, E., Chesani, F., Poluzzi, A., Capuzzo, D., 2022. Statistical stability indices for lime: Obtaining reliable explanations for machine learning models. Journal of the Operational Research Society 73, 91–101.

[54] Wong, P.K., Zhong, J., Yang, Z., Vong, C.M., 2016. Sparse bayesian extreme learning committee machine for engine simultaneous fault diagnosis. Neurocomputing 174, 331–343.

[55] Xue, B., Zhang, M., Browne, W.N., Yao, X., 2015. A survey on evolutionary computation approaches to feature selection. IEEE Transactions on evolutionary computation 20, 606–626.

[56] Zhao, X., 2015. Lab test of three fault detection and diagnostic methods' capability of diagnosing multiple simultaneous faults in chillers. Energy and buildings 94, 43–51.

[57] Zhu, M., Xia, J., Jin, X., Yan, M., Cai, G., Yan, J., Ning, G., 2018. Class weights random forest algorithm for processing class imbalanced medical data. IEEE access 6, 4641–4652.

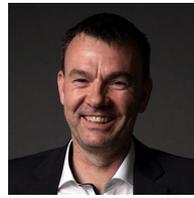
Andreas rausch is the director of the Institute for Software and Systems Engineering at the TU Clausthal. From 2007 to 2018, he was the head of the chair for Software Systems Engineering at the TU Clausthal. Until early 2007, he was the head of the chair for Software Architecture at the University of Kaiserslautern. In 2001, he obtained his doctorate from the University of Munich under Prof. Dr. Manfred Broy. His research in the field of software engineering focuses on software architecture, model-based software engineering, and process models, with more than 140 publications worldwide. He is involved in the management of and participation in a number of international, European, and national research projects.

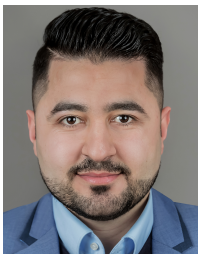
Mohammad Abboush received the Ph.D. degree from the Clausthal University of Technology, in 2024. He received the M.Sc. degree in mechatronic engineering from Siegen University, Siegen, Germany, in 2018. Since 2024, he has been a Research Associate with the Institute for Software and System Engineering, Clausthal University of Technology, Clausthal-Zellerfeld, Germany, where he is currently pursuing the PostDoc degree. His current research interests include the verification and validation of automotive software systems, real-time hardware-in-the-loop (HIL) simulation, intelligent fault detection and diagnosis, fault injection test, reliability and safety analysis, and machine learning. His work focuses on the development of intelligent failure analysis methods based on machine learning approach for real-time validation of automotive software systems.

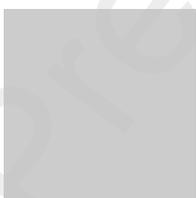
Ehab Ghannoum received his bachelor's degree in Computer Engineering from Qatar University. He later obtained his master's degree in Computer Science from the TU Clausthal, Germany, where he is currently pursuing his Ph.D. studies at Institute for Software and System Engineering, TU Clausthal, Germany.